\documentclass[multphys,vecphys]{svmult}

\newcommand{\lesssim}{\mathrel{\raise2pt\hbox{\rlap{\hbox{\lower4pt\hbox{$
\sim$}}}\hbox{$<$}}}}
\newcommand{\gtrsim}{\mathrel{\raise2pt\hbox{\rlap{\hbox{\lower4pt\hbox{$
\sim$}}}\hbox{$>$}}}}

\usepackage{makeidx}         
\usepackage{graphicx}        
\usepackage{multicol}        
\usepackage[bottom]{footmisc}

\usepackage{wrapfig}

\makeindex             

\begin{document}

\title*{How can we make sure we detect dark matter?}  
\titlerunning{How can we make sure we detect dark matter?}
\author{Paolo Gondolo}
\institute{Department of Physics, University of Utah, 115 S 1400 E Rm
  201, Salt Lake City, UT 84112-0830 \texttt{paolo@physics.utah.edu}}
%
%
\maketitle

\begin{abstract}
More and more claims of having detected WIMP dark matter are being put forward. Some are discussed here, stressing the importance of exploiting distinctive signatures to ascertain their WIMP origin. The best signals for WIMP discovery are characterized by special features that make them recognizable as due to WIMPs and nothing else. Sometimes, however, a single feature, although accountable for in theoretical models, may not be enough to make sure that we have detected WIMPs. This is because the theory of WIMPs and their distribution in the galaxy is still very uncertain, and allows for many possibilities. What are needed are {\it experimental} verifications of the claimed signals, either by discovering {\it unmistakable} features, or by detecting several kinds of signals that can all be explained by the {\it same} WIMP model.
\end{abstract}

One of the most intriguing results to come out of recent cosmological
observations is that about 90\% of the mass of the Universe is not
made of protons, neutrons, electrons, or any other known particle, but
of something unknown that does not shine. Discovering the composition
of this so-called non-baryonic dark matter is one of the big
challenges of modern physics and cosmology.

Proposals as to the nature of non-baryonic dark matter do not lack. Is
it made of axions (neutral particles suggested to explain the
smallness of CP violation in the strong interactions)? Or is it made
of WIMPs (weakly interacting massive particles that arise naturally in
extensions of the Standard Model of particle physics, such as
supersymmetry)? Or is non-baryonic dark matter made of something else,
or a combination of all that?

Following the tradition of experimental science, the way to find out
the nature of non-baryonic dark matter is to detect its constituents,
either directly by recording their collisions with a detector, or
indirectly by observing products of their reactions in planets, stars,
or galaxies.

The last ten years have seen more and more claims of having detected
dark matter in the form of weakly interacting massive particles (WIMPs). Three of these claims will be described below: (1) a
distinctive signal variation in a direct detection experiment, (2)
high-energy gamma-rays from the center of our Galaxy, and (3) an
excessive flux of positrons in cosmic rays.

However, explanations that do not invoke WIMPs exist for gamma-rays
and positrons, and other direct detection experiments have not
observed any signal from WIMPs (although straightforward comparisons
are difficult).

So, has WIMP dark matter been detected? What is the real origin of the
detected signals? Or more proactively: how can we make sure we detect
WIMP dark matter?

This question has been asked repeatedly in the past, and
several methods have been proposed to distinguish a dark matter signal
from an ordinary one. In recent years, however, an excessive reliance on theory has interfered with an open-minded but critical interpretation of the experimental results. Current theories of WIMPs still leave a lot of possibilities as to their particle properties (mass, couplings, etc.) and astrophysical characteristics (density distribution, velocity distribution, etc.). Rather than theoretical arguments, what are needed are {\it experimental} verifications of the claimed signals, either by discovering {\it unmistakable} features which can only be explained by the presence of WIMPs, or by detecting several kinds of WIMP signals that can all be explained by the {\it same} theoretical model of WIMPs.

\section{Non-baryonic cold dark matter}

The existence of non-baryonic dark matter is supported by varied
cosmological measurements. Of great relevance are the values of the
matter and energy densities of the Universe at the present time. These
densities can be determined by means of several cosmological data: the
temperature fluctuations in the cosmic microwave background (CMB), the
distance-luminosity relation for supernovas, the distribution of
galaxies on large scales, the abundance of light elements (primordial
nucleosynthesis), etc. The density values so obtained are compatible
with all current astrophysical and cosmological observations, from the
internal motions of galaxies and galaxy clusters to studies of weak
gravitational lensing.  Ref.~\cite{Spergel} finds the following values
\cite{Bennett} for the current matter and energy densities, ${\rm
  \Omega} h^2$, in units of $ 1.879 \times 10^{-29} $ g/cm$^3$ (i.e.\ 
18.79 yg/m$^3$ or 1.689 nJ/m$^3c^2$):
\begin{itemize}
\item[$\bullet$] \hskip-0.5em a negligible density in relativistic
  particles (``radiation''; e.g., the CMB photons contribute only
  ${\rm\Omega}_\gamma h^2 = (2.467\pm0.004) \times 10^{-5}$);
  \vspace{0.5\baselineskip}
\item[$\bullet$] \hskip-0.5em ${\rm\Omega_\Lambda} h^2 = 0.36 \pm 0.04
  $ in a smoothly distributed component (dark energy);
  \vspace{0.5\baselineskip}
\item[$\bullet$] \hskip-0.5em $ {\rm\Omega_m} h^2 =
  0.135^{+0.008}_{-0.009}$ in non-relativistic particles (``matter''),
  of which \vspace{0.3\baselineskip}
\begin{itemize}
\item[\hskip-0.5em --] \hskip-1em $ {\rm\Omega_b} h^2 = 0.0224 \pm
  0.0009$ in protons and neutrons (baryons),
\item[\hskip-0.5em --] \hskip-1em $ {\rm\Omega_{HDM}} h^2 < 0.0076 $
  (95\% CL) in non-baryonic hot dark matter, \vspace{0.3\baselineskip}
\item[\hskip-0.5em --] \hskip-1em $ {\rm\Omega_{CDM}} h^2 =
  0.113^{+0.008}_{-0.009}$ in non-baryonic cold dark matter.
\end{itemize}
\end{itemize}
It is the excess of total matter density ($\simeq 0.135$) over
baryonic matter density ($\simeq 0.0224$) that constitutes the
evidence for non-baryonic dark matter.

None of the known elementary particles can account for non-baryonic
dark cold matter. The obvious Standard Model candidates would be
neutrinos, but the measurements of the neutrino mass squared
differences, ${\rm\Delta}(m^2) \lesssim 10^{-3}$ eV$^2$, and the
experimental upper bound of 3~eV on the mass of the neutrino produced
in tritium beta decay, impose that the masses of all three known
neutrinos are $m_{\nu} < 3$~eV. Neutrinos so light constitute {\it
  hot} dark matter, and have to be included in $ {\rm\Omega_{HDM}}
h^2< 0.0076 $ (95\% CL). Thus no known particle is a candidate for
non-baryonic cold dark matter.

Scores of hypothetical particles have been proposed as cold dark
matter candidates over the past several decades. They range from new
particles in well-founded extensions of the Standard Model, to
possible particles inspired by recent theoretical ideas. To the first
category belong  an extra heavy neutrino, the axion, and
the lightest supersymmetric particle (the neutralino, the gravitino,
or the sneutrino). In the second category are particles like
WIMPZILLAs, solitons (B-balls and Q-balls), self-interacting dark
matter, string-inspired dark matter, Kaluza-Klein dark matter, etc.

\section{Dark matter WIMPs and their detection}

A class of non-baryonic dark matter candidates is of interest here
because of several recent claims to their detection: weakly
interacting massive particles, or WIMPs. WIMPs are appealing because
of a simple mechanism that can produce the observed value of their
cosmic density. Assume that in the early Universe WIMPs were in
thermal and chemical equilibrium with the rest of the matter and
radiation. As the Universe expanded and cooled down, the chemical
reactions coupling WIMPs to the rest of the world slowed down and
eventually stopped, leaving a constant number of WIMPs in a volume that
expands with the Universe. Numerically, the correct present-time
density of WIMPs is obtained for matter-WIMP couplings of the order of
electroweak couplings, and WIMP masses in the 1 GeV--100
TeV range. These characteristics give these particles their name.
Examples of WIMPs are a heavy neutrino and the lightest neutralino. The latter arises in supersymmetric extension of the Standard Model, and is one of the most popular candidates for non-baryonic dark matter.

Signals from dark matter WIMPs can be either direct or indirect.
Direct signals are due to collisions of dark matter WIMPs with nuclei
in a detector. A very sensitive low-background detector records the amount of energy deposited by WIMPs in collisions with nuclei, and in the future also the direction of motion of the struck nucleus.

Indirect signals are due to the products of WIMP
reactions in planets, stars, or galaxies. The most common reaction is WIMP annihilation: WIMPs can annihilate with anti-WIMPs, if present, or with themselves, if, like the neutralino, they are their own anti-particle. Out of the products of WIMP annihilation, neutrinos, positrons, anti-protons, and high-energy gamma-rays are those of most interest because they are rarely produced by usual astrophysical processes. WIMP annihilations occur at a detectable rate where WIMPs are concentrated, as in the center of the Sun, the center of the Earth, and the inner regions of galactic halos, ours in particular. Neutrino telescopes, gamma-ray telescopes, and cosmic ray detectors are used to search for WIMPs indirectly.

The next three sections discuss signals that can or have been attributed to WIMPs, stressing the importance of exploiting distinctive signatures to ascertain their WIMP origin.

\section{The HEAT positron excess}

In two separate balloon flights with different detectors, the
HEAT collaboration \cite{HEAT} has observed more cosmic ray positrons above $\sim$7 GeV than
expected in current models of cosmic ray propagation in our
galaxy. In these models, positrons arise as secondary particles in the interactions of primary cosmic rays with interstellar matter. Modifications of these models could in principle account for the extra positrons (and similar extra photons observed in EGRET \cite{EGRET}), but no proposed modification can yet reproduce all observable cosmic ray data (see discussion in~\cite{MS04}, e.g.). WIMP annihilation can \begin{wrapfigure}[19]{r}[0pt]{0.49\textwidth}
\vskip-2\baselineskip
\centering
\includegraphics[width=0.49\textwidth]{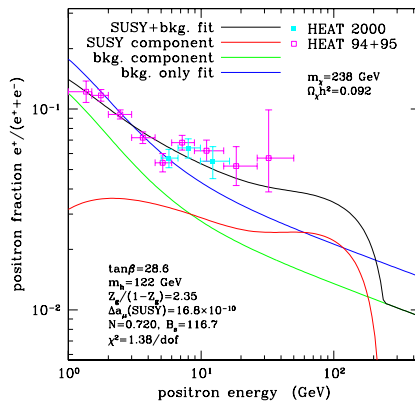}
\caption{The HEAT positron excess can be interpreted as due to annihilation of WIMPs, here a 238 GeV neutralino in the minimal supersymmetric standard model (from~\protect\cite{BEFG01}).}
\label{fig:HEAT}
\end{wrapfigure}  
also be invoked to explain the extra positrons, as illustrated in figure~\ref{fig:HEAT}.

Do the positron data distinguish between the suggested origins for the excess? Not by themselves, in that they lack a clear and unique signature that they are due to WIMPs. The positron spectrum predicted by WIMP
annihilation lacks any discriminating feature, with the exception of a reduction in flux and then a cut-off as the energy increases towards the WIMP mass. Any such flux reduction can however be pushed beyond any foreseeable maximum detectable energy by simply raising the WIMP mass.
In the absence of a distinguishing feature from WIMPs, it is hard to draw conclusions on the origin of the positron excess.

\section{Gamma-rays from the Galactic Center}

In principle, gamma-rays from WIMP annihilation offer a characteristic signature in their spectrum: a gamma-ray line~\cite{gamma-line}. The line originates in the annihilation of WIMPs into a pair of photons, each photon carrying an energy equal to the  WIMP mass, between 10 GeV and 100 TeV. No astrophysical process is known to produce a gamma-ray line at these energies. This makes the WIMP gamma-ray line an ideal signature for WIMPs.  
\begin{wrapfigure}[13]{r}{0.48\textwidth}
\vskip-2\baselineskip\centering
\includegraphics[width=0.48\textwidth]{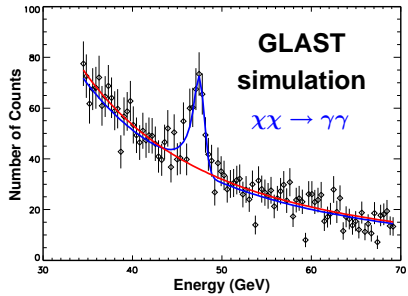}
\vskip-0.5\baselineskip
\caption{Simulation showing the GLAST capability of detecting a gamma-ray line from WIMP annihilation (from~\protect\cite{GLAST}).}
\label{fig:GLAST}
\end{wrapfigure}
\indent Searches for the WIMP gamma-ray line are continuing, but no line has been detected yet. The challenge is twofold: both a large number of photons and a fine energy resolution are needed. Figure~\ref{fig:GLAST} shows that the space-born gamma-ray telescope GLAST, scheduled for launch in 2006, is expected to have such capabilities.

In the meantime, another source of gamma-rays from WIMPs has been used to claim their detection: the gamma-ray continuum. These are gamma-rays generated in the decay of secondary products, such as pions, produced by WIMP annihilation into quarks, W, or Z bosons. Contrary to the gamma-ray line, continuum gammas from WIMPs lack a characteristic feature, except for a flux reduction and cut-off near the WIMP mass. In this respect, they are similar to cosmic ray positrons from WIMPs. Raising the WIMP mass pushes the flux reduction beyond the observable energies.

\begin{figure}[t]
\begin{minipage}{0.49\textwidth}
\centering
\includegraphics[width=0.95\textwidth]{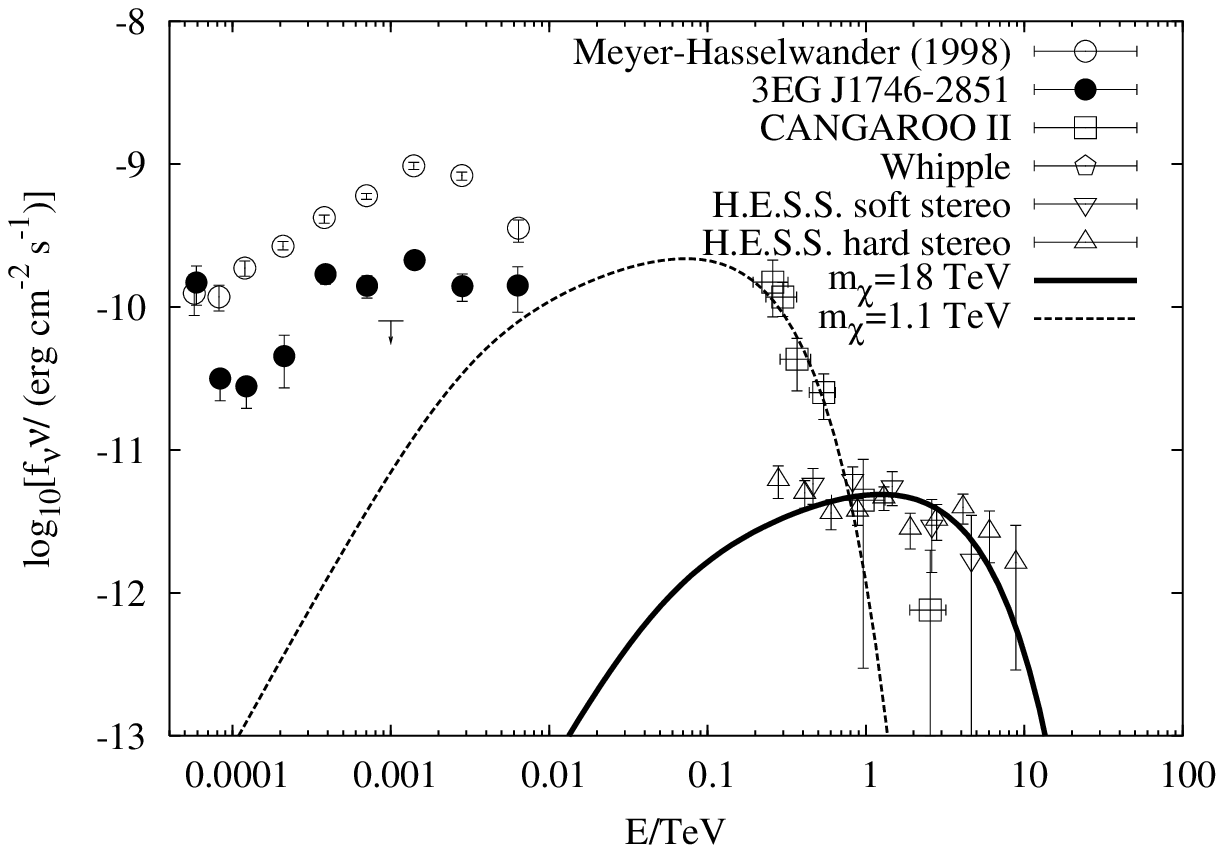}
\caption{Both CANGAROO and HESS Galactic Center data can be fitted to WIMP annihilation spectra, although with different WIMP masses (from~\protect\cite{Horns}).}
\label{fig:Horns}
\end{minipage}
\hfill
\begin{minipage}{0.49\textwidth}
\includegraphics[width=1.0\textwidth]{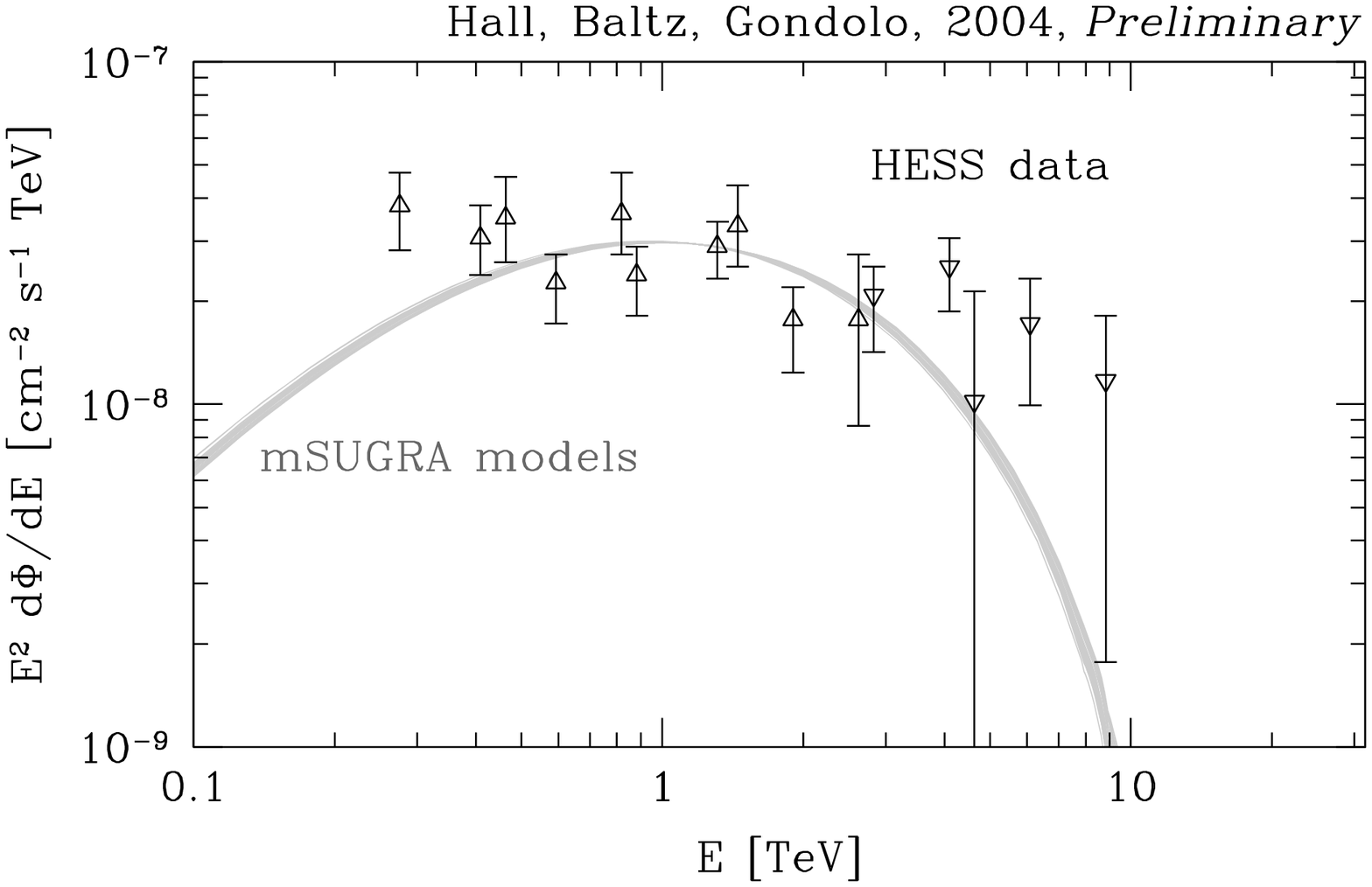}
\caption{The HESS spectrum can be fitted even to neutralinos in the minimal supergravity model (from~\protect\cite{HBG}).}
\label{fig:HBG}
\end{minipage}
\end{figure}

The possibilities are limitless. For example, the early-2004 CANGAROO \begin{wrapfigure}[16]{l}{0.4\textwidth}
\vskip-1.5\baselineskip
\centering
\includegraphics[width=0.4\textwidth]{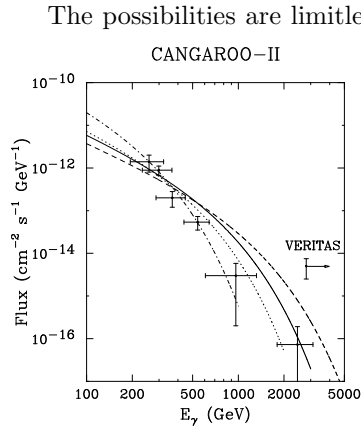}
\vskip-0.5\baselineskip
\caption{WIMP annihilation can fit the CANGAROO gamma-ray data from the Galactic Center (from~\protect\cite{Hooper}).}
\label{fig:Hooper}
\end{wrapfigure}
report of high-energy gamma-rays from the Galactic Center~\cite{CANGAROO} has been interpreted as due to annihilations of $\sim$ 1 TeV WIMPs (see figure~\ref{fig:Hooper}). The CANGAROO data can also be explained by appropriate modeling of accretion flows around the black hole at the Galactic Center~\cite{CANG-astro}. The mid-2004 HESS observation of gamma-rays from the same region~\cite{HESS} has a very different spectrum from CANGAROO's, but it can also be interpreted either with appropriate (but different) accretion flows~\cite{CANG-astro} or as due to WIMP annihilation, this time with $\sim$ 20 TeV WIMPs (figure~\ref{fig:Horns}). If this mass seems too high to supersymmetry aficionados, it may be amusing to see that even minimal supergravity models allow for $\sim$ 10 TeV neutralinos compatible with cosmological and astrophysical bounds and the HESS data (figure~\ref{fig:HBG}).

The lack of an unmistakable WIMP signature in the gamma-ray continuum makes it unsuitable as a primary indicator of the presence of WIMPs. A gamma-ray line would be desirable.

\section{The DAMA annual modulation}

An excellent signature for WIMPs in direct detection has been known for years: the annual modulation~\cite{ann.mod.}. The Earth motion periodically changes the relative speed of Earth and WIMPs, causing the WIMP flux on Earth, and thus the WIMP detection rate, to vary in time and to repeat itself once every year. The details of the annual pattern depend on the WIMP velocity distribution. For example, the date of maximum rate is set by the most common arrival direction of the WIMPs, and happens in June for the canonical halo model with Maxwellian velocity distribution, but may occur in December for Sikivie's cold infall model (see~\cite{GG00}). Similarly, the amplitude of the modulation depends on the halo model.

The DAMA collaboration has observed an annual modulation in their \begin{wrapfigure}[10]{l}{0.55\textwidth}
\vskip-2.5\baselineskip
\centering
\includegraphics[width=0.55\textwidth]{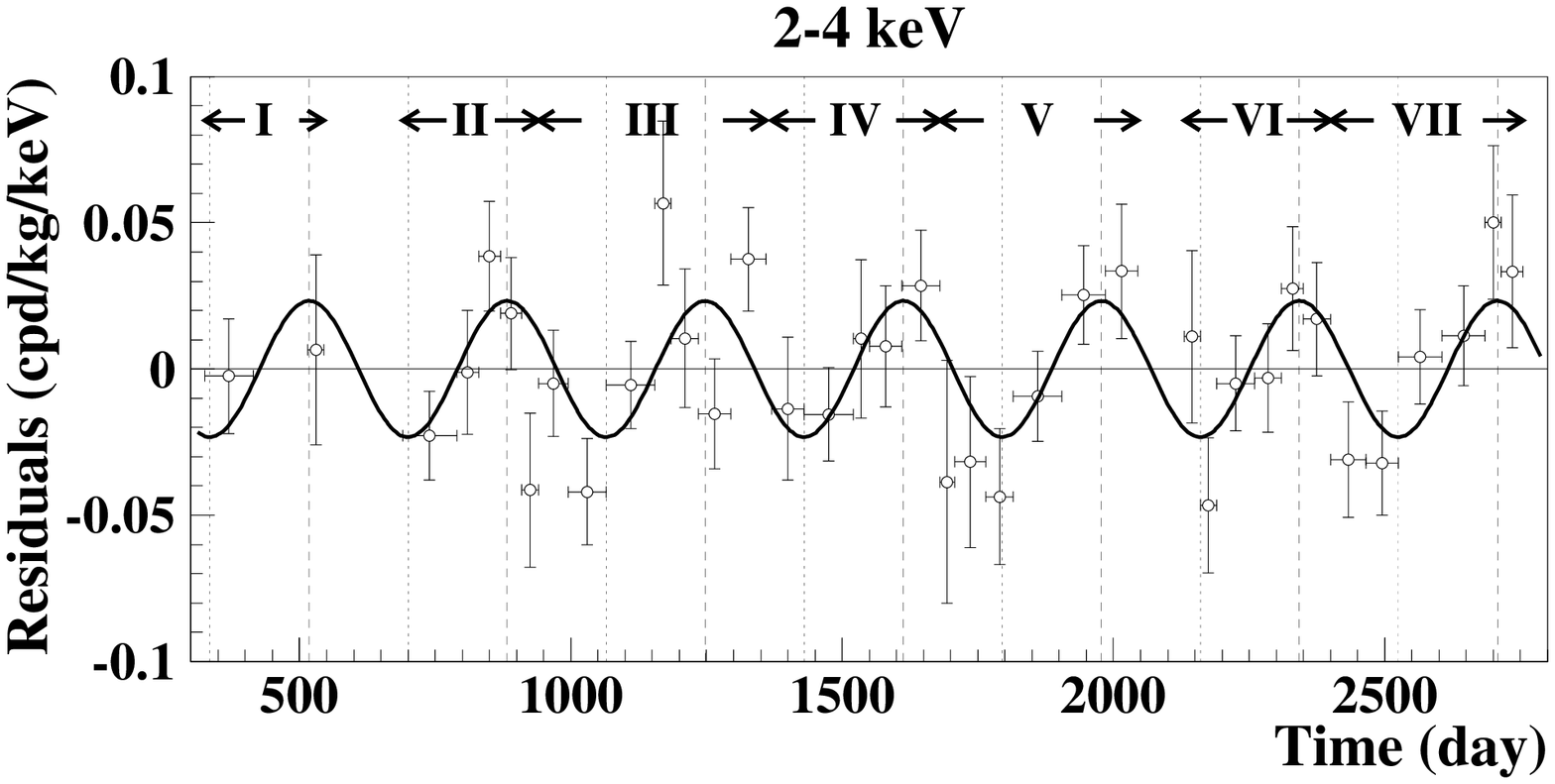}
\vskip-\baselineskip
\caption{The DAMA annual modulation (from~\protect\cite{DAMA}).}
\label{fig:DAMA}
\end{wrapfigure}
sodium iodide data (figure~\ref{fig:DAMA}) and has attributed it to WIMPs. No valid alternative explanation has been put forward yet, but no WIMP signal has been observed in any other direct detection experiment either. However, comparison of the various experimental results, which are obtained with different targets, is marred by the need of uncertain theoretical assumptions about the WIMP mass, interaction, and halo model. In fact, the expected event rate depends on the product of the WIMP-nucleus cross section and the WIMP flux on Earth. The cross section scales differently with the atomic mass of the target nucleus according as the WIMP interacts with the nuclear mass or the nuclear spin. The flux depends on the distribution of WIMP velocities, which is probably more complicated than an arbitrarily-assumed Maxwellian. Indeed, current hierarchical models of galaxy formation entail halo substructure and streams of dark matter. (Further details on the direct detection rate can be found in~\cite{cargese}.)

\begin{figure}[t]
\includegraphics[width=0.45\textwidth]{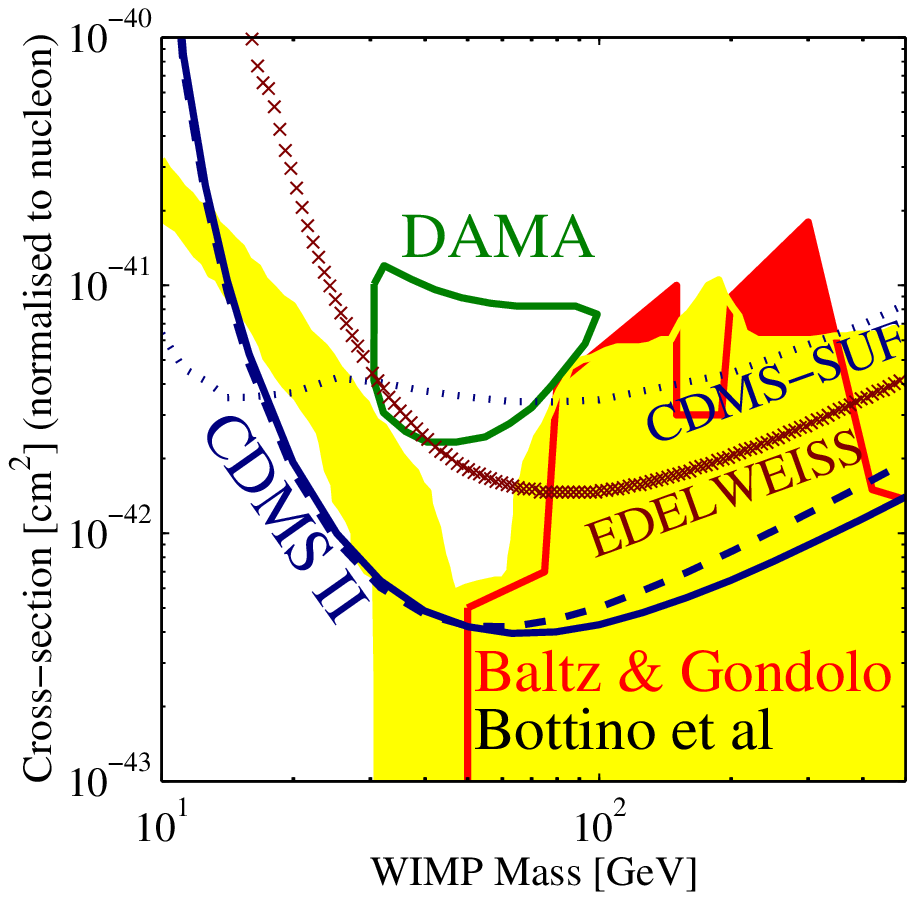}
\lower6pt\hbox{\includegraphics[width=0.47\textwidth]{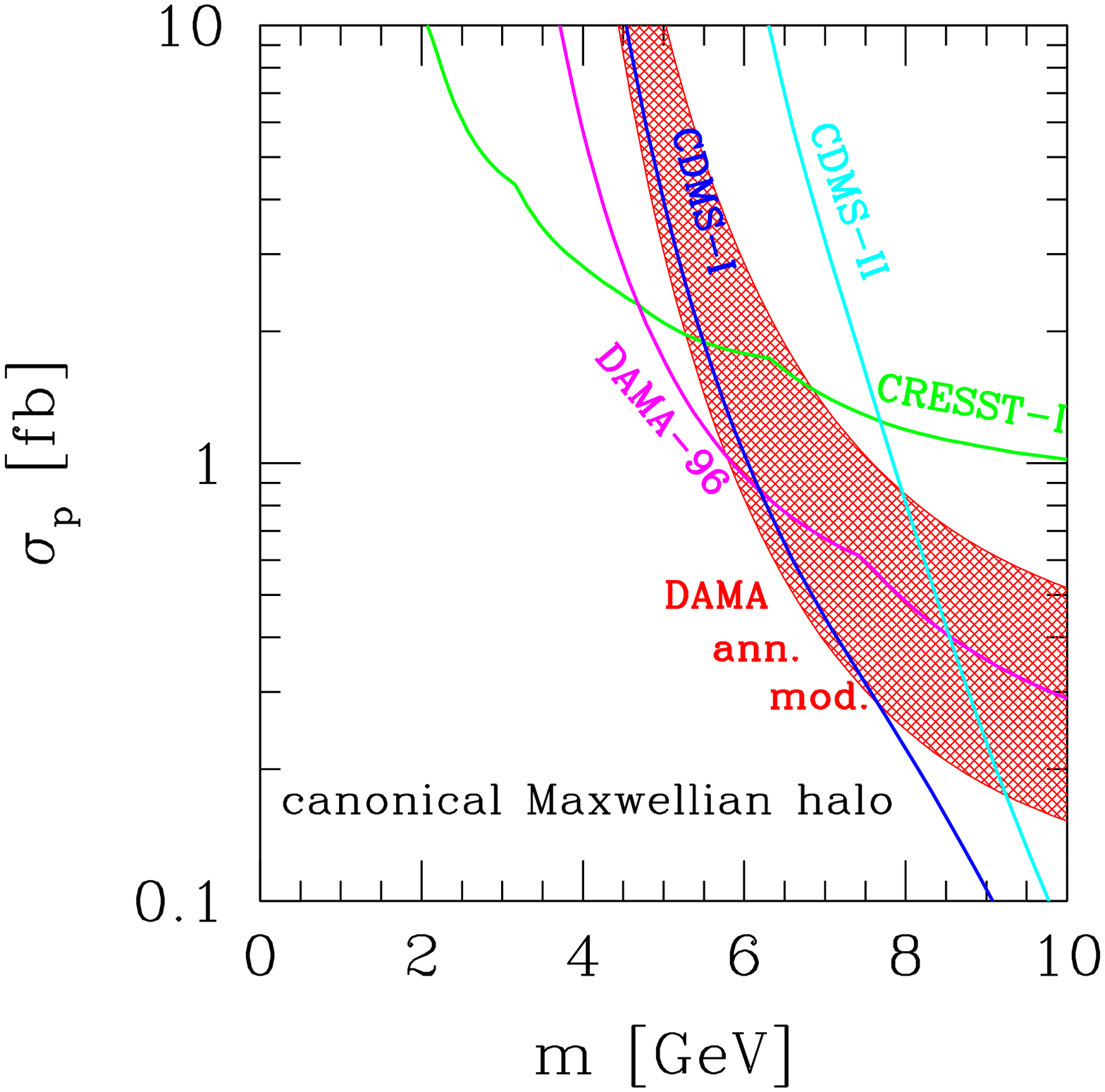}}
\vskip-0.5\baselineskip
\caption{{\it Left}: for spin-independent WIMP-nucleus interactions and a canonical halo with Maxwellian velocity distribution, CDMS~\protect\cite{CDMS} excludes the original DAMA region, which was however artificially cut at a WIMP mass of 30 GeV~\protect\cite{DAMA1}. {\it Right}: relaxing the cut, the DAMA region is still (narrowly) allowed at lower masses~\protect\cite{ggDAMA}.}
\vskip-0.5\baselineskip
\label{fig:CDMS}
\end{figure}

Neglect of the theoretical assumptions involved has given rise to many controversies. For example, the first DAMA analysis~\cite{DAMA1} was artificially restricted to WIMPs heavier than 30 GeV, on the basis of some theoretical prejudice on the nature of the WIMPs. Under the assumption of a spin-independent cross section, the latest CDMS data~\cite{CDMS} exclude the original DAMA region (figure~\ref{fig:CDMS}a), but relaxing the artificial restriction on the WIMP mass there remains the possibility of a (now admittedly narrow) region at low WIMP masses (figure~\ref{fig:CDMS}b). The moral is: do not use theoretical prejudices when analyzing data.

\begin{wrapfigure}[26]{l}{0.55\textwidth}
\hskip-1em\includegraphics[width=0.6\textwidth]{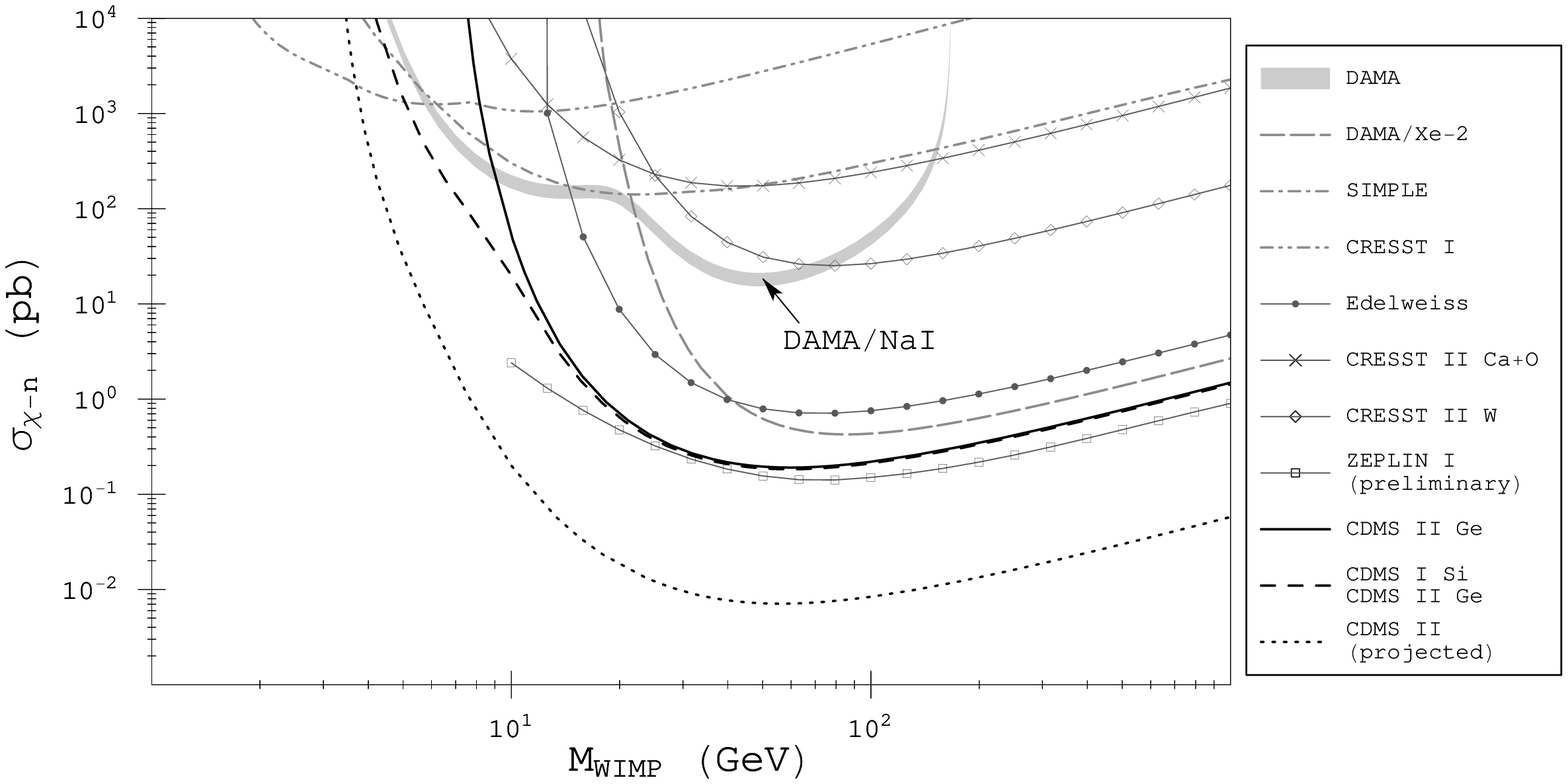}
\vskip-0.01\baselineskip
\hskip-1em\includegraphics[width=0.6\textwidth]{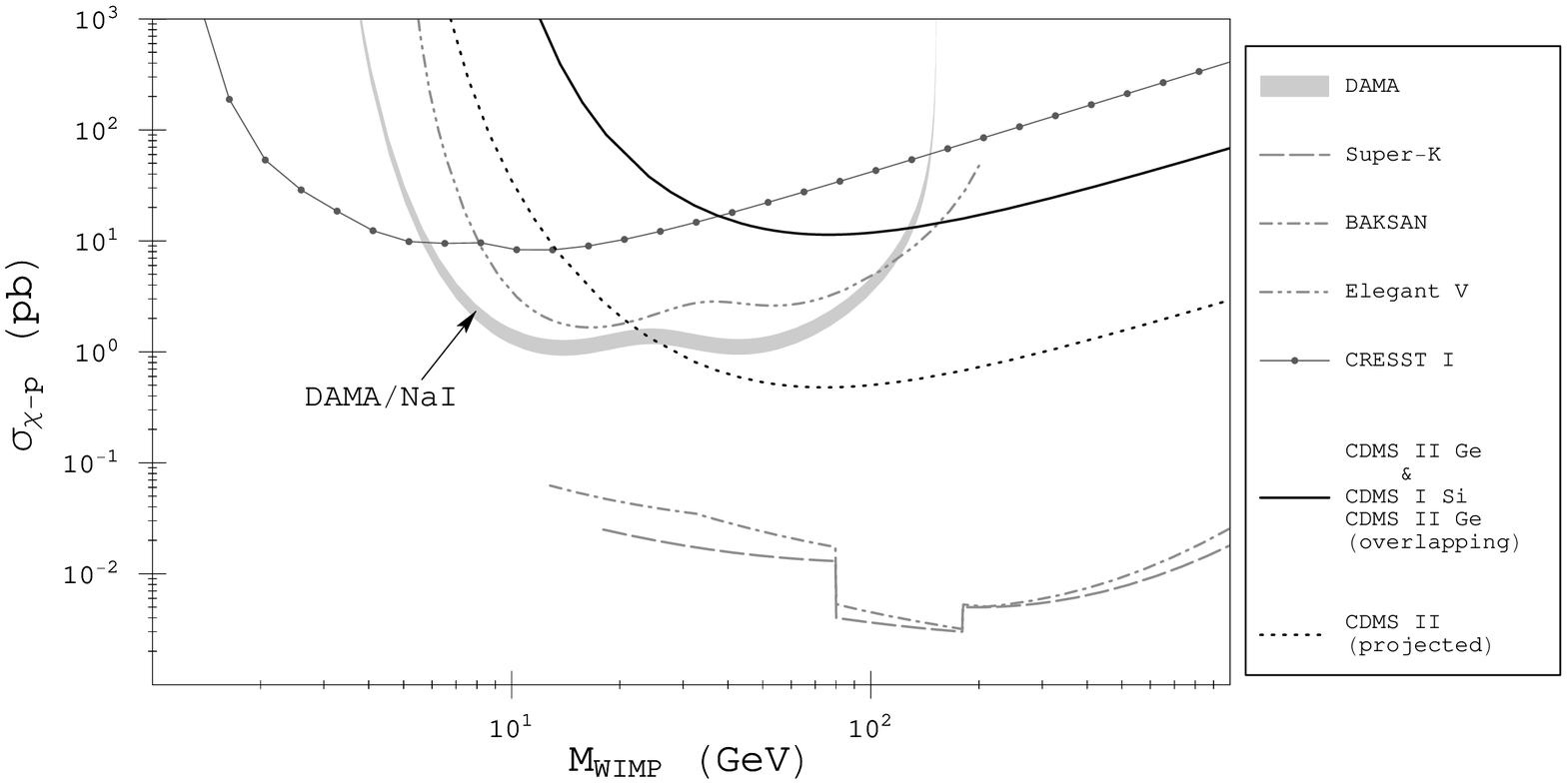}
\vskip-0.25\baselineskip
\caption{{\it Upper panel}: if WIMPs interact predominantly with neutrons, existing data exclude the DAMA region for spin-dependent interactions and a canonical Maxwellian halo. {\it Lower panel}: if WIMPs interact predominantly with protons, the DAMA region may still be allowed under appropriate assumptions on WIMP annihilation in the Sun (from~\cite{SGF}).}
\label{fig:SGF}
\end{wrapfigure}

Another example in which the theoretical assumptions play a subtle role is the case of spin-dependent cross sections. For a canonical Maxwellian halo, if WIMPs interact predominantly with neutrons, existing data exclude the DAMA region, while if WIMPs interact predominantly with protons, the DAMA region may still be allowed, but only under further assumptions. As shown in figure~\ref{fig:SGF}, the most stringent limit on WIMP-proton spin-dependent interactions comes from the absence of indirect neutrino signals 
from WIMP annihilation in the Sun. This places an upper limit of $\sim$10 GeV on the WIMP mass (and a further analysis of lower energy neutrinos may 
lower this limit even further). However, there may be no anti-WIMPs in the Sun with which the WIMPs can annihilate, and as a consequence there may be no neutrino bounds on the spin-dependent WIMP-proton cross section. This would make the DAMA region allowed up to WIMP masses of $\sim$100 GeV. Notice that the neutralino, a fashionable candidate, is its own anti-particle, and thus for it the neutrino bounds cannot be avoided; restricting attention to the neutralino is however a theoretical prejudice.

In the face of all these difficulties of interpretation, other WIMP signatures in direct detection would be helpful. Very promising are detectors that can record not only the energy deposited by the WIMP but also the direction of motion of the nucleus after the collision. One such detector, DRIFT, is currently under construction~\cite{DRIFT}. It will be possible to use the recoil direction of the nuclei to discriminate a WIMP signal from background, for instance more WIMPs should come from the direction of motion of the Solar System than from the opposite direction. 

\begin{figure}[t]
\begin{minipage}{0.49\textwidth}
\centering
\includegraphics[width=\textwidth]{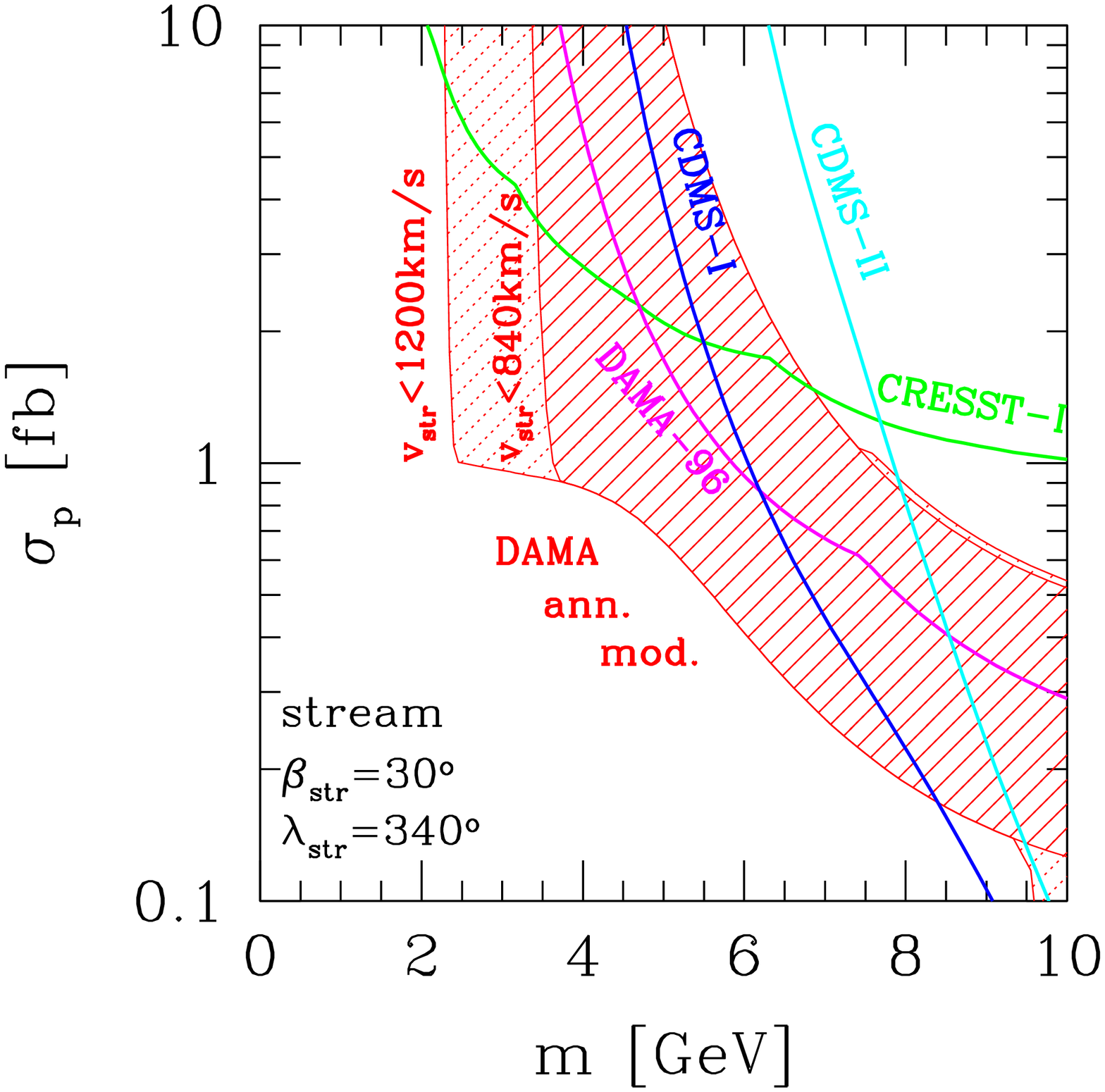}
\caption{Dark matter streams enlarge the possibilities for the DAMA region (from~\cite{ggDAMA}).}
\label{fig:ggDAMA}
\end{minipage}
\hfill
\begin{minipage}{0.49\textwidth}
\vskip6pt
\includegraphics[width=0.97\textwidth]{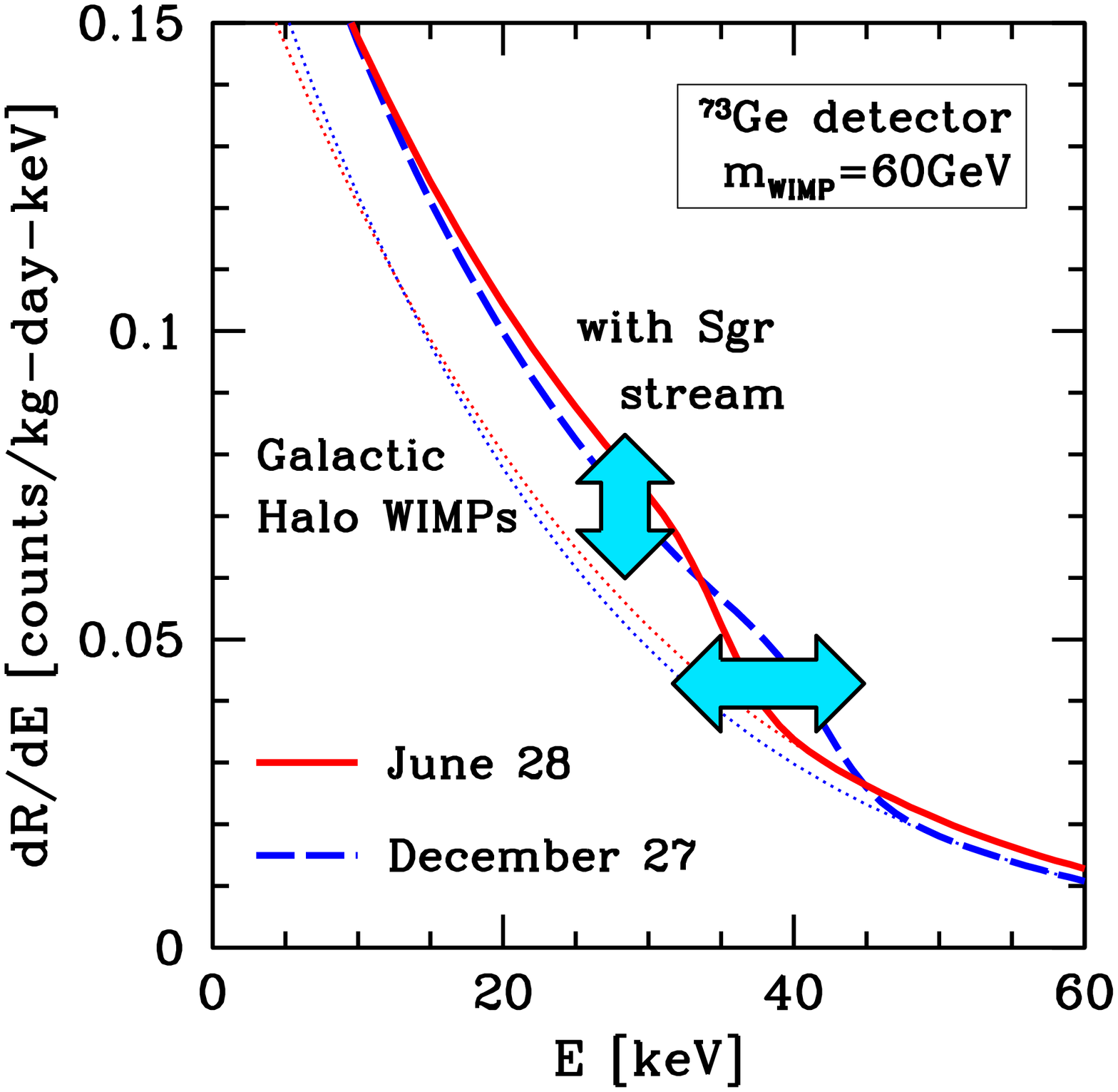}
\caption{A new combined modulation of both rate and end-point energy may be a powerful signature of WIMPs in a dark matter stream (from~\protect\cite{FGN}).}
\label{fig:FGN}
\end{minipage}
\end{figure}

Another signature for WIMPs has been proposed in~\cite{FGN} in case a stream of dark matter passes through the Solar System. Streams are common in simulations of galaxy formation, and have already been observed in our galaxy. One of them in particular, the stream associated with the tidal disruption of the Sagittarius dwarf galaxy, may pass by the Solar System. Streams through the Solar System may even enlarge the possibilities for the DAMA region, provided they come from roughly ahead of the Solar System motion (see figure~\ref{fig:ggDAMA}).
The new signature proposed in~\cite{FGN} is a combined annual modulation of the rate and of the highest energy that WIMPs in the stream can impart to a nucleus. This highest energy shows as a step-like feature in the nucleus recoil spectrum, and the location of the step is predicted to vary in energy with a period of one year (figure~\ref{fig:FGN}). No background spectrum is expected to behave in this way.

With additional experimental signatures for WIMPs in direct detection we may be able to understand the origin of a claimed signal.

\section{How can we make sure we detect dark matter?}

The best signals for WIMP discovery are characterized by special features that make them recognizable as due to WIMPs and nothing else. For example, gamma-rays from WIMP annihilation should show a gamma-ray line in correspondence to the WIMP mass; high-energy ($\gtrsim$GeV) neutrinos  from the Sun or the Earth cannot be produced by anything else but WIMPs; the direct detection rate, and for WIMPs in a stream, the highest recoil energy,  should follow a modulation with a period of a year; etc. It is these features one should look for in searching for dark matter WIMPs. 

History has shown however (e.g.\ for the DAMA annual modulation) that a single feature, although accountable for in theoretical models, may not be enough to make sure that we have detected WIMPs. This is because the theory of WIMPs and their distribution is still very uncertain, and allows for many possibilities.
What is needed is {\it experimental} confirmation from a variety of detectors, and perhaps from different kinds of signals that can be explained within the {\it same} WIMP model. Different kinds of signals will in any case be needed to determine all the interesting WIMP properties, such as mass, cross section, density, etc. Finally, to be really convinced, we will probably have to produce WIMPs in the laboratory, perhaps with high energy colliders.

\end{document}